\documentclass[useAMS,usenatbib]{mn2e}
\usepackage{graphicx}
\usepackage{amsmath}
\usepackage{url}
\usepackage{xspace}
\usepackage[breaklinks,colorlinks,citecolor=blue]{hyperref}

\newcommand{\spose}[1]{\hbox to 0pt{#1\hss}}
\newcommand{\approxpropto}{\mathrel{\spose{\lower 3pt\hbox{$\sim$}}
	\raise 2.0pt\hbox{$\propto$}}}
\def\approxgt{\mathrel{\spose{\lower 3pt\hbox{$\sim$}}
	\raise 2.0pt\hbox{$>$}}}
\def\approxlt{\mathrel{\spose{\lower 3pt\hbox{$\sim$}}
	\raise 2.0pt\hbox{$<$}}}

\def\beq{\begin{equation}}
\def\eeq{\end{equation}}
\def\omm{\Omega_{\rm m}} 
\def\omb{\Omega_{\rm b}}\def\fb{f_{\rm b}}
\def\dndlnM{\mathrm{d}n/\mathrm{d}\ln M_{500}}
\def\hmsun{h^{-1} \mathrm{M}_\odot}

\newcommand{\GO}{GO\xspace}
\newcommand{\PC}{PC\xspace}
\newcommand{\FO}{FO\xspace}
\newcommand{\DMone}{DM1\xspace}
\newcommand{\DMtwo}{DM2\xspace}

\title[Baryons \& the cluster mass function]
{Impact of baryons on the cluster mass function and cosmological parameter determination}
\author[S. J. Cusworth et al.]
  {Sam J. Cusworth,$^1$\thanks{Email: cusworth@jb.man.ac.uk}
  Scott T. Kay,$^1$ Richard A. Battye,$^1$ and Peter A. Thomas$^{2}$ \\
  $^1$Jodrell Bank Centre for Astrophysics, School of Physics and Astronomy, The University of Manchester, Manchester M13 9PL\\
  $^{2}$Astronomy Centre, Department of Physics and Astronomy, University of Sussex, Brighton BN1 9QH
  }

\begin{document}

\date{Accepted 2014 January 15. Received 2014 January 06; in original form 2013 September 15}

\pagerange{\pageref{firstpage}--\pageref{lastpage}} \pubyear{2013}

\label{firstpage}

\maketitle

\begin{abstract}
Recent results by the Planck collaboration have shown that cosmological parameters derived from the cosmic microwave background anisotropies and cluster number counts are in tension, with the latter preferring lower values of the matter density parameter, $\omm$, and power spectrum amplitude, $\sigma_8$. 
Motivated by this, we investigate the extent to which the tension may be ameliorated once the effect of baryonic depletion on the cluster mass function is taken into account. 
We use the large-volume Millennium Gas simulations in our study, including one where the gas is pre-heated at high redshift and one where the gas is heated by stars and active galactic nuclei (in the latter, the self-gravity of the baryons and radiative cooling are omitted). 
In both cases, the cluster baryon fractions are in reasonably good agreement with the data at low redshift, showing significant depletion of baryons with respect to the cosmic mean. 
As a result, it is found that the cluster abundance in these simulations is around 15 per cent lower than the commonly-adopted fit to dark matter simulations by Tinker et al (2008) for the mass range $10^{14}-10^{14.5}\hmsun$. 
Ignoring this effect produces a significant artificial shift in cosmological parameters which can be expressed as $\Delta[\sigma_8(\omm/0.27)^{0.38}]\simeq -0.03$ at $z=0.17$ (the median redshift of the \textit{Planck} cluster sample) for the feedback model. 
While this shift is not sufficient to fully explain the \textit{Planck} discrepancy, it is clear that such an effect cannot be ignored in future precision measurements of cosmological parameters with clusters. 
Finally, we outline a simple, model-independent procedure that attempts to correct for the effect of baryonic depletion and show that it works if the baryon-dark matter back-reaction is negligible.

\end{abstract}


\section{Introduction}\label{sec:intro}
Clusters of galaxies map peaks in the cosmic density field and as such can be used to determine information about the Universe on the largest scales~\citep{Voit2005a,Allen2011,Kravtsov2012}.
In particular, the abundance of clusters as a function of mass and redshift has been shown to be a particularly sensitive probe of cosmological parameters $\omm$ and $\sigma_8$, the matter density parameter and the linear rms matter fluctuation within a spherical top-hat of 8 $h^{-1}$ Mpc radius respectively\footnote{Throughout we express the Hubble parameter today as $H_0=100h \mathrm{~km~s}^{-1}\mathrm{ Mpc}^{-1}$.}~\citep{Vikhlinin2009,Rozo2010,Reichardt2012,HasselfieldMatthew2013,PlanckCollaboration2013}.
In order to link the observed mass function of clusters to an underlying cosmology one must appeal to an analytic description of cluster abundance~\citep{Press1974,Bond1991,Sheth2002} or to one of many numerical studies investigating dark matter halo formation, e.g.\ \citet{Jenkins2001,Tinker2008,Watson2012}\footnote{See \citet{Murray2013} for a recent comparison of mass functions in the literature.}.
One of the most commonly adopted descriptions of cluster halo abundance is the~\citet{Tinker2008} mass function (TMF hereafter).

The implicit assumption made in linking simulated dark matter halo masses with galaxy cluster masses is that the ratio of baryons to dark matter within clusters does not differ significantly from the cosmic value.
This assumption, however, has been challenged by multi-wavelength observations~\citep{Lin2003,Giodini2009,Lagana2011}.
It has also been shown in $N$-body simulations that the pressure forces within baryonic gas are capable of segregating the distribution of collisional gas relative to pressure-less dark matter~\citep{NavarroJ.F.1993} thereby changing the baryon fraction within clusters~\citep{Crain2007}.
Numerical studies have also shown that galaxy formation processes and non-gravitational heating can modify the baryon fraction~\citep{mccarthy2011,Planelles2013} and thereby the total mass within clusters~\citep{Stanek:2008am}.

The measurement of the cluster mass function from observations requires the calibration of an \textit{observable-mass} ($X-M$) relation, where common observables, $X$, are X-ray luminosity, galaxy richness and Sunyaev-Zel'dovich (SZ) flux.
Multiple ongoing observational surveys including the \textit{Planck} mission ~\citep{PlanckCollaboration2013}, the South Pole Telescope survey (SPT;~\citealt{Reichardt2012}), Dark Energy Survey (DES;~\citealt{Collaboration2005}) and the \textit{XMM} Cluster Survey (XCS;~\citealt{Romer2001,Sahlen2009}) have made the cosmological analysis of the galaxy cluster mass function one of their key scientific goals.
In order to parameterise the systematic uncertainties in the measurement of a given scaling relation, {e.g.} of incorrectly assuming hydrostatic equilibrium in clusters, the mass bias parameter $b_{\rm hyd}=1-M_{\rm true}/M_{X}$ is commonly employed, where $M_{X}$ is the mass inferred from observable $X$.

In the near future, large volume observational surveys such as \textit{eROSITA}~\citep{Pillepich2012}, \textit{Euclid}~\citep{LaureijsR.2011a}, the Large Synoptic Survey Telescope (LSST;~\citealt{LSSTScienceCollaboration2009}) and the proposed PRISM mission\footnote{\url{http://www.prism-mission.org/}} will detect a greater number of galaxy clusters than ever before.
It is therefore of great importance that the cluster mass function is accurately calibrated against theoretical predictions~\citep{Reed2012}.

Recent results from the \textit{Planck} cluster survey~\citep{PlanckCollaboration2013} have been found to be in tension with cosmological parameter determinations made using anisotropies in the \textit{Cosmic Microwave Background} (CMB;~\citealt{PlanckCollaboration2013a}).
It has been argued that the discrepancy between the two measured values of $\sigma_8$ and $\omm$ could be due, in part, to cluster biases and selection effects.
Alternatively, it has been proposed that the influence of additional physical processes, such as the influence of massive neutrinos on the power spectrum, could lead to an underestimation in the mass function.

In this paper, we use large cosmological simulations with baryonic physics to investigate whether such tension can at least in part be explained by the effects of baryonic depletion in clusters. 
Such an effect, due to gas being expelled by feedback processes, produces a shift in the cluster mass function to lower abundance at fixed mass which if not accounted for in the cosmological analysis, leads to derived values for cosmological parameters ($\omm$ and $\sigma_8$) that are systematically underestimated.

The remainder of the paper can be summarised as follows. 
In Section~\ref{sec:Sim_selection}, we outline details of the simulations used and how the cluster samples were defined. 
In Section~\ref{sec:cluster_MF}, our main results are presented, quantifying the effect of the baryon depletion on the mass function and its subsequent effect on the cosmological parameters $\omm$ and $\sigma_8$, before suggesting a simple corrective procedure. 
Finally, in Section~\ref{sec:discussion} we discuss our results in the context of other work in the literature and draw conclusions.

\section{Simulations and halo selection}\label{sec:Sim_selection}
\subsection{Millennium Gas simulations}
We use results from three cosmological Millennium Gas simulations (MGS;~\citealt{Hartley:2007dm,Stanek:2008am,Stanek2010,short2010,Young2011,Kay:2011cg}) and two dark matter-only versions of the same volumes.
The MGS are designed to include the dynamics of gas that were not present in the dark matter-only Millennium simulations.
Each simulation in the suite is run with a different treatment of large scale baryonic physics.
We group these simulations into two ``generations'', determined by the underlying cosmological model employed.

\subsubsection{First generation: \GO \& \PC models}
In the \textit{Gravitation Only} (\GO) simulation, first described in~\citet{Crain2007}, baryonic gas is only permitted to change in entropy through shock heating.
As a counterpoint to the adiabatic \GO simulation, in the \textit{Pre-heating \& Cooling} (\PC) simulation, described in~\citet{Hartley:2007dm}, radiative cooling of gas was implemented (assuming a metalicity $Z=0.3Z_\odot$).
Furthermore, in order to emulate the effects of high redshift galaxy formation and reproduce the observed X-ray luminosity-temperature relation at $z\simeq 0$, the gas within the volume was uniformly heated to 200 keV cm$^2$ at $z=4$.

The \GO and \PC simulations were run using the \texttt{Gadget-2} code \citep{Springel2005a} with the same cosmological model as the Millennium simulation~\citep{Springel2005}.
The parameters used were $\omm=0.25$, $\omb=0.045$, $h=0.73$ and $\sigma_8=0.9$; consistent with the first year \textit{Wilkinson Microwave Anisotropy Probe} results (\textit{WMAP}1;~\citealt{Spergel2003}).
Because of computational constraints, the simulations were run with a slightly decreased mass resolution compared to the original Millennium run.
A downgraded version of the Millennium initial conditions was used in the \GO and \PC simulations.
At early times ($z>3$) the gravitational softening length was fixed in comoving coordinates to $\epsilon=100 h^{-1}\mathrm{kpc}$, whereas at late times ($z<3$) the softening was fixed in physical coordinates to $\epsilon=25 h^{-1}\mathrm{kpc}$.
The particle masses were set to $m_{\rm dm}=1.4\times 10^{10}\hmsun$ and $m_{\rm gas}=3.1 \times 10^9 \hmsun$ for the dark matter and gas respectively.
In both simulations the dark matter was evolved self-consistently with the gas.
As such the baryons influence the formation and growth rate of dark matter structures.

We compare the first generation MGS to a version of the original Millennium simulation~\citep{Springel2005} with the same initial conditions, mass resolution and gravitational softening lengths as the \GO and \PC models.
We will refer to this simulation as \DMone.

\subsubsection{Second generation: \FO model}
In the \textit{Feedback Only} (\FO) simulation, the effects of stochastic Active Galactic Nuclei (AGN) and supernovae feedback on the gas dynamics were inferred using the semi-analytic model of~\citet{Guo2011}.
For full details regarding the treatment of the gas dynamics in \FO see~\citet{Short:2012ck}\footnote{The MGS2-FO simulation described in~\citet{Hilton2012} implemented the same physical model, albeit with a smaller simulated volume.}.
The principle improvement of the \FO simulation over the \PC is that the baryonic feedback is better physically motivated, although radiative cooling is not included.
One caveat of note is that in the \FO model the baryonic contribution to the gravitational potential is ignored. In other words, the gas is evolved with zero gravitational mass and so there is no baryon-dark matter back-reaction.

The \FO simulation was run using an updated version of the Gadget code, \texttt{Gadget-3} at the resolution of the original Millennium simulation.
Smaller softening lengths than the first generation simulations ($\epsilon=37 h^{-1}\mathrm{kpc}$ in comoving coordinates before $z\simeq3$ and $\epsilon=9.3 h^{-1}\mathrm{kpc}$ in physical coordinates thereafter) were set.
The masses of dark matter and gas particles were set to $7.8\times 10^8 \hmsun$ and $3.1\times 10^8 \hmsun$ respectively.

We will compare \FO to an updated version of the dark matter-only Millennium simulation (\DMtwo;~\citealt{Springel2005}). 
Both simulations in the second generation used the same set of initial conditions, generated using second order Lagrangian perturbation theory (2LPT;~\citealt{Scoccimarro1998}).
In addition, the \FO and \DMtwo simulations were carried out using cosmological parameters consistent with the 7-year \textit{WMAP} results (\textit{WMAP}7;~\citealt{Komatsu:2010fb}): $\omm=0.272$, $\omb=0.0455$, $h=0.704$ and $\sigma_8=0.81$.

\subsection{Cluster sample}
Clusters were identified from the simulated density field using combinations of the friends of friends (FOF;~\citealt{Davis1985a}), \texttt{SUBFIND}~\citep{Springel2001} and spherical overdensity~\citep{Press1974,Lacey1994} algorithms (see \citet{Knebe2011} for a review of halo-finding techniques). 
In all of the analysis presented here we consider clusters with mass $\mathcal{O}\left( 10^{14} \hmsun\right)$.
These clusters correspond to groups containing $>10^4$ particles.

In the first generation MGS (\GO, \PC) and \DMone simulation, clusters were identified using the procedure outlined in~\citet{Kay:2011cg}.
Briefly, the dark matter particles were initially grouped using a FOF algorithm, with dimensionless linking length $b=0.1$.
The linking length parameter was chosen to be smaller than the canonical $b=0.2$ in order to avoid the so called ``over-bridging'' problem whereby distinct, neighbouring haloes are linked together \citep{Knebe2011}.
Next, the centre of each cluster was identified as the dark matter particle with the most negative gravitational potential energy.
Finally the bulk properties of the clusters (mass, radii, etc.) were calculated using the properties of all particles within spherical regions of overdensity $\Delta={\rho\left(<R_{\Delta}\right) / \rho_{\rm c}\left(z\right)}$ where $\rho_{\rm c}(z)=(3 H_0^2/8\pi G) E(z)^2$ is the cosmic critical density at redshift $z$ and $E(z)^2=\omm (1+z)^3 +\Omega_\Lambda$.
Throughout we take $\Delta=500$, therefore cluster masses are defined
\beq
M_{500}=500{4 \pi \over 3} \, R_{500}^3 \, \rho_{\rm c}(z),
\eeq
where $R_{500}$ is the proper radius of the spherical overdensity.
For reasons described in Section~\ref{sec:cl_MF}, we will consider clusters with $M_{500}=10^{14}-10^{14.5}\hmsun$. In the \GO, \PC and \DMone simulations there are 1016, 800 and 965 clusters respectively, at $z=0$.

In the second generation MGS (\FO) and dark matter-only \DMtwo simulation, a similar procedure was implemented.
First, a FOF algorithm was run with $b=0.2$.
In order to avoid the over-bridging problem, \texttt{SUBFIND} was used to identify gravitationally bound structures within each FOF group.
We then took the centre of the most massive substructure within the FOF groups to be the cluster centre.
Finally, the bulk properties were calculated using the spherical overdensity algorithm.
At $z=0$ there are 707 and 830 clusters with $M_{500}$ within the range of interest, for the \FO and \DMtwo models, respectively. 
The different number of clusters found in \DMone and \DMtwo is mainly due to the fact that they utilise different cosmological models.

It should be noted that since the peak of the density field within a FOF group is considered to be the centre of a cluster in the analysis of both generations, the procedures used here are largely equivalent.
The selection criteria in both generations exclude low mass clusters whose centres lie within the $R_{500}$ of more massive clusters, in line with other studies~\citep{Tinker2008}.

\subsection{Baryon fraction}\label{sec:bary_dep}
\begin{figure}
\centering
\includegraphics[width=8.5cm]{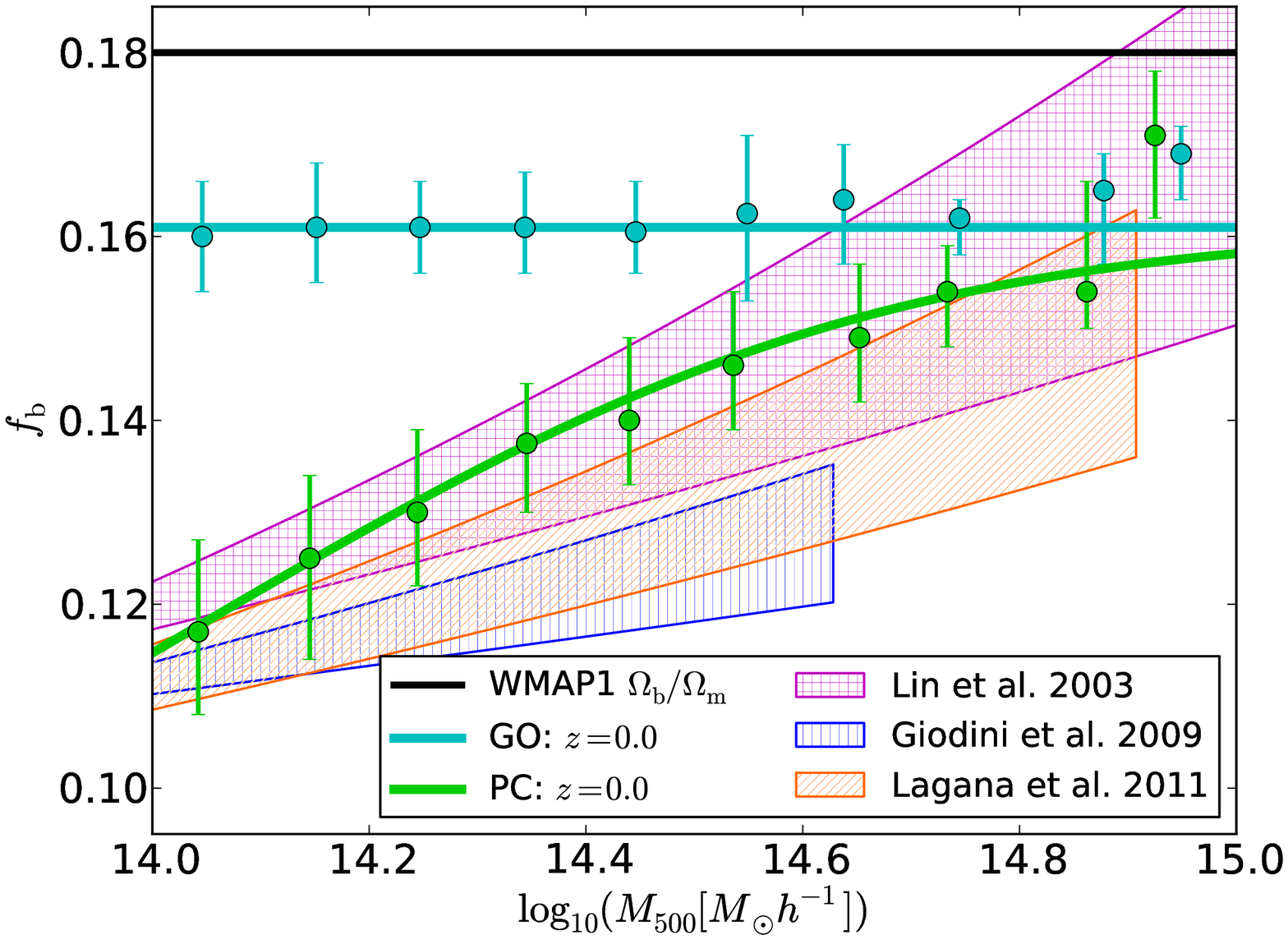}
\includegraphics[width=8.5cm]{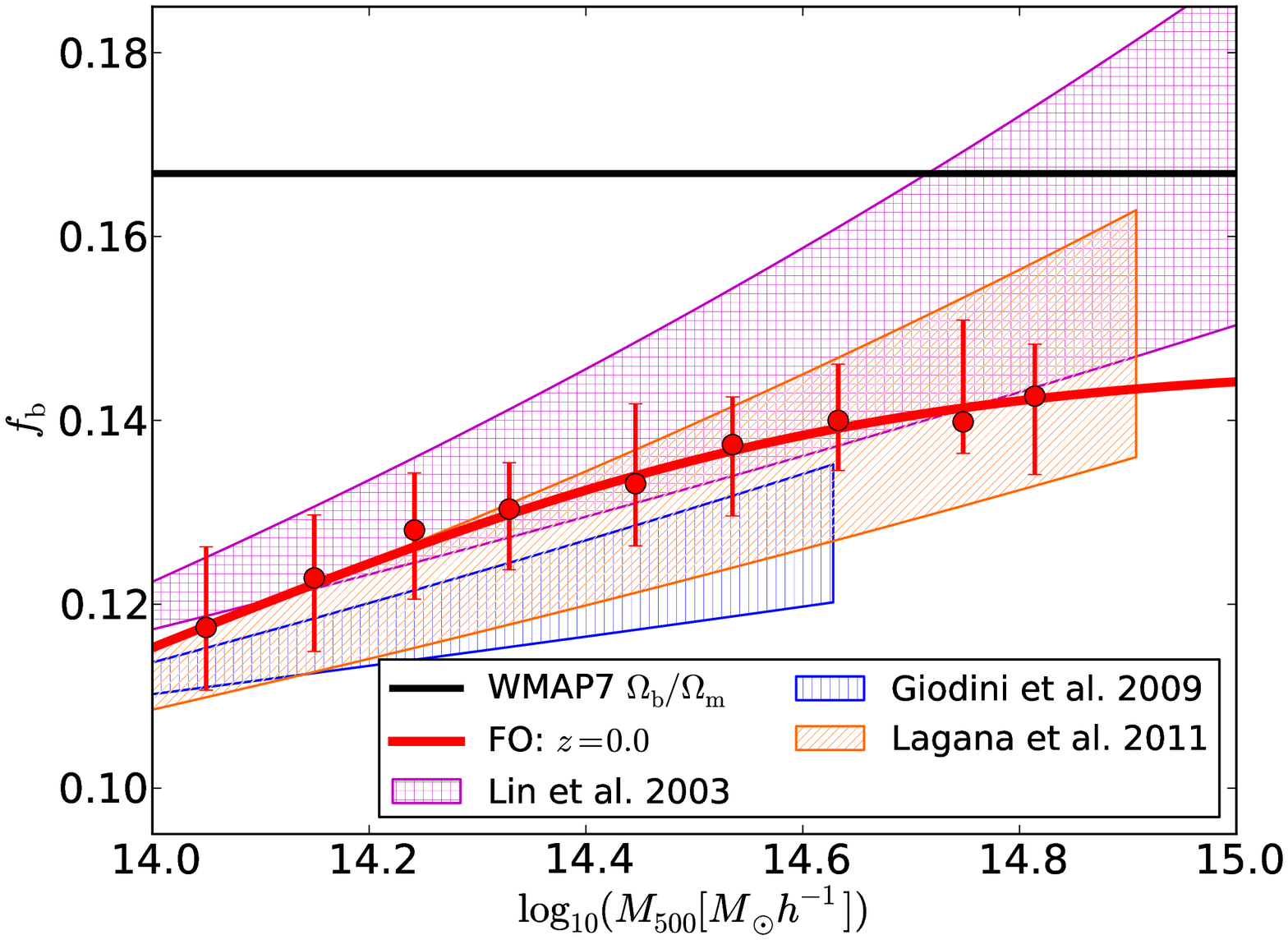}
\caption{The baryon fraction as a function of $M_{500}$ is shown for both generations of MGS. results from \GO and \PC are show in the top panel (cyan and green respectively) while results from \FO are shown in red in the lower panel. The simulated data derived from simulation outputs at $z=0$ are plotted, where bars indicate the 16th and 84th percentiles of the distribution and the coloured points show the median value within each mass bin. In the top panel we plot the fits to the simulated data (\GO and \PC) from~\citet{Young2011} and in the lower panel we plot the fit computed for the \FO simulation. We also plot the low redshift observational bounds of~\citet{Lin2003},~\citet{Giodini2009} and~\citet{Lagana2011} in the orange, purple and blue regions respectively. In each panel the cosmic mean, $\omb/\omm$, calculated using the corresponding \textit{WMAP}1/7 parameters is also shown in black.}
\label{fig:baryon_fraction}
\end{figure}
In the simulated clusters we define the baryon fraction
\beq
\fb=\frac{M_{*}\left(< R_{500}\right)+M_{\rm gas}\left(<R_{500}\right)}{M_{500}},
\eeq
where $M_{*}$ and $M_{\rm gas}$ are the masses of stars and gas respectively within $R_{500}$.

The baryon fractions calculated from the clusters in the two generations of MGS are plotted in Fig.~\ref{fig:baryon_fraction} as a function of cluster mass $M_{500}$.
The baryon faction for the \GO and \PC clusters, presented in~\citet{Young2011}, is also shown in the top panel of Fig.~\ref{fig:baryon_fraction} (cyan and green curves respectively).
We follow~\citet{Young2011} and fit the \FO baryon fraction scaling relation to the function
\beq
\log_{10} \fb = \log_{10} f_0 + s\left[\mu -\frac{1}{4} \ln \left(1+\exp(4\mu) \right) \right]
\eeq
where $\mu=\log_{10}\left(M/M_{\rm piv} \right)$ is the mass variable scaled by a pivot mass $M_{\rm piv}$ and $f_0$ and $s$ are two free parameters.
The best fitting parameters, for $\log_{10}\left(M_{\rm piv} [\hmsun]\right)=14.47$, were $f_0=0.146\pm 0.001 $ and $s=0.204\pm 0.008$.
Errors quoted here were calculated using bootstrap resampling, keeping $M_{\rm piv}$ fixed.

In both panels of Fig.~\ref{fig:baryon_fraction} we also plot the best fits to the observational data of~\citet{Lin2003},~\citet{Giodini2009} and~\citet{Lagana2011} along with the associated uncertainties.
Estimations of $\fb$ within clusters require knowledge of bulk properties such as mass and radius, measurements of the intra-cluster gas and observations of the stellar mass distribution.
These observations are often made difficult by the contributions of intra-cluster light and fainter dwarf galaxies.

In non-radiative simulations such as \GO, the baryonic gas distribution within a cluster becomes more extended than the dark matter because it is able to gain energy in halo merger events \citep{Crain2007}. 
The resulting baryon fraction within $R_{500}$ is therefore slightly reduced relative to the cosmic mean in a manner that is independent of halo mass.
As a counterpoint, the baryon fraction in the \PC and \FO simulations is scale dependent. 
In the \PC case, gas heated within a small halo is more likely to be ejected than gas in a halo with a deeper gravitational potential well.
Similarly, the net effect of AGN and supernovae feedback is to eject more baryonic gas from lower mass clusters.

It is clear from Fig.~\ref{fig:baryon_fraction} that the baryon fraction in $z\simeq 0$ galaxy clusters, both observed and simulated, is less than the cosmic mean $\omm/\omb$ for the mass range plotted.
There is excellent agreement between the simulated baryon fraction and the observational bounds plotted in both the \PC and \FO models\footnote{We note that the baryon fraction in low mass simulated clusters ($M_{500}\approxlt 10^{14}\hmsun$; not shown) was found to be significantly lower than the cosmic mean ($\fb\simeq 0.1$) in agreement with~\citet{Lin2003} and simulations which include AGN feedback~(e.g.~\citealt{mccarthy2011,Planelles2013}.}.
While the baryon fraction within clusters can itself be used as a cosmological probe~\citep{Allen2011}, here we use it as a test of the validity of the gas physics model employed in our simulations.

The baryonic depletion in the hydrodynamically simulated clusters leads directly to lower values of $M_{500}$ relative to the dark matter-only counterparts.
The effect of the depletion on the cluster mass function and subsequent cosmological parameter estimations is the subject of the following section.

\section{Results}\label{sec:cluster_MF}
\subsection{Cluster mass function}\label{sec:cl_MF}
\begin{figure*}
\centering
\includegraphics[width=8.5cm]{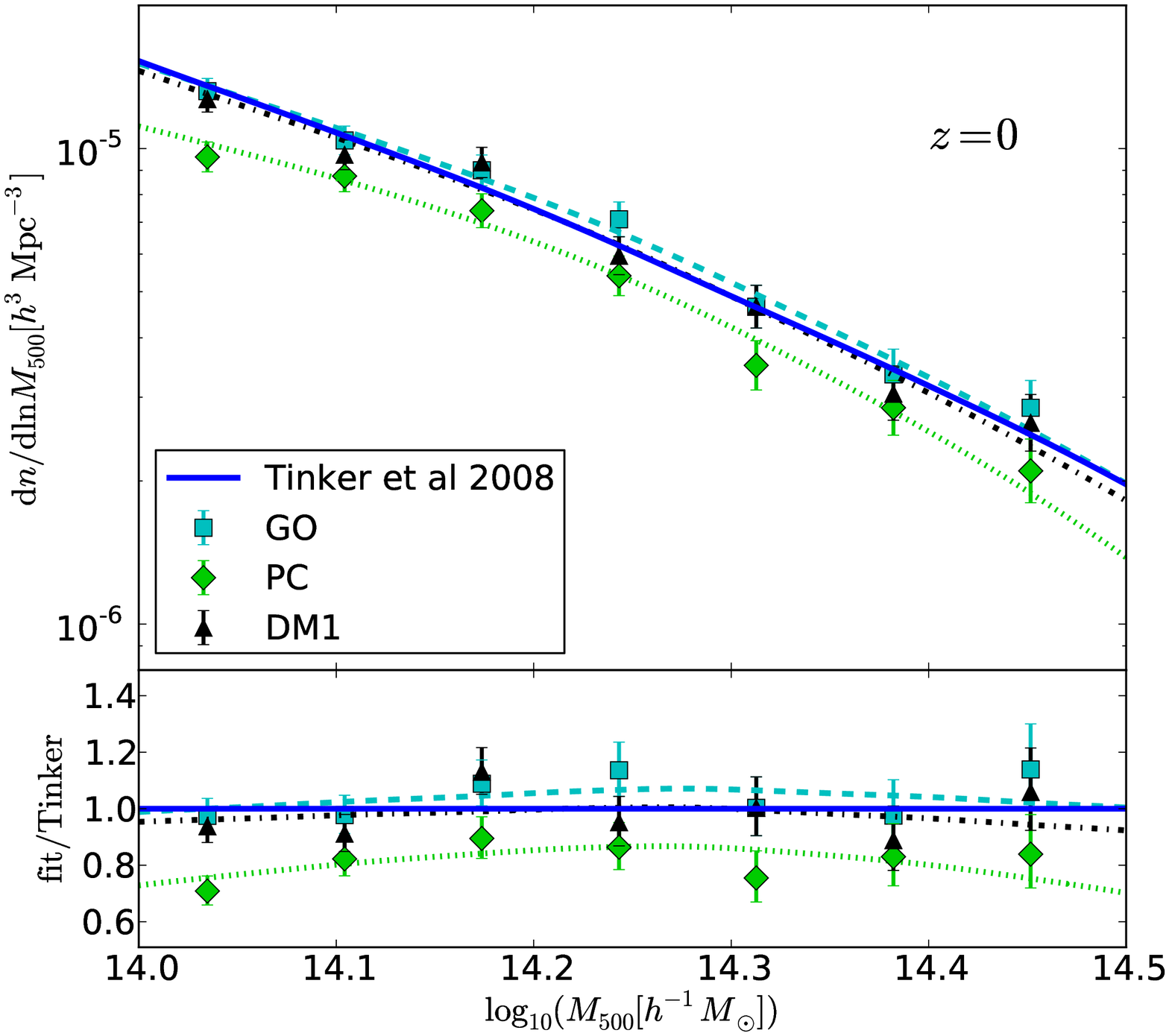}
\includegraphics[width=8.5cm]{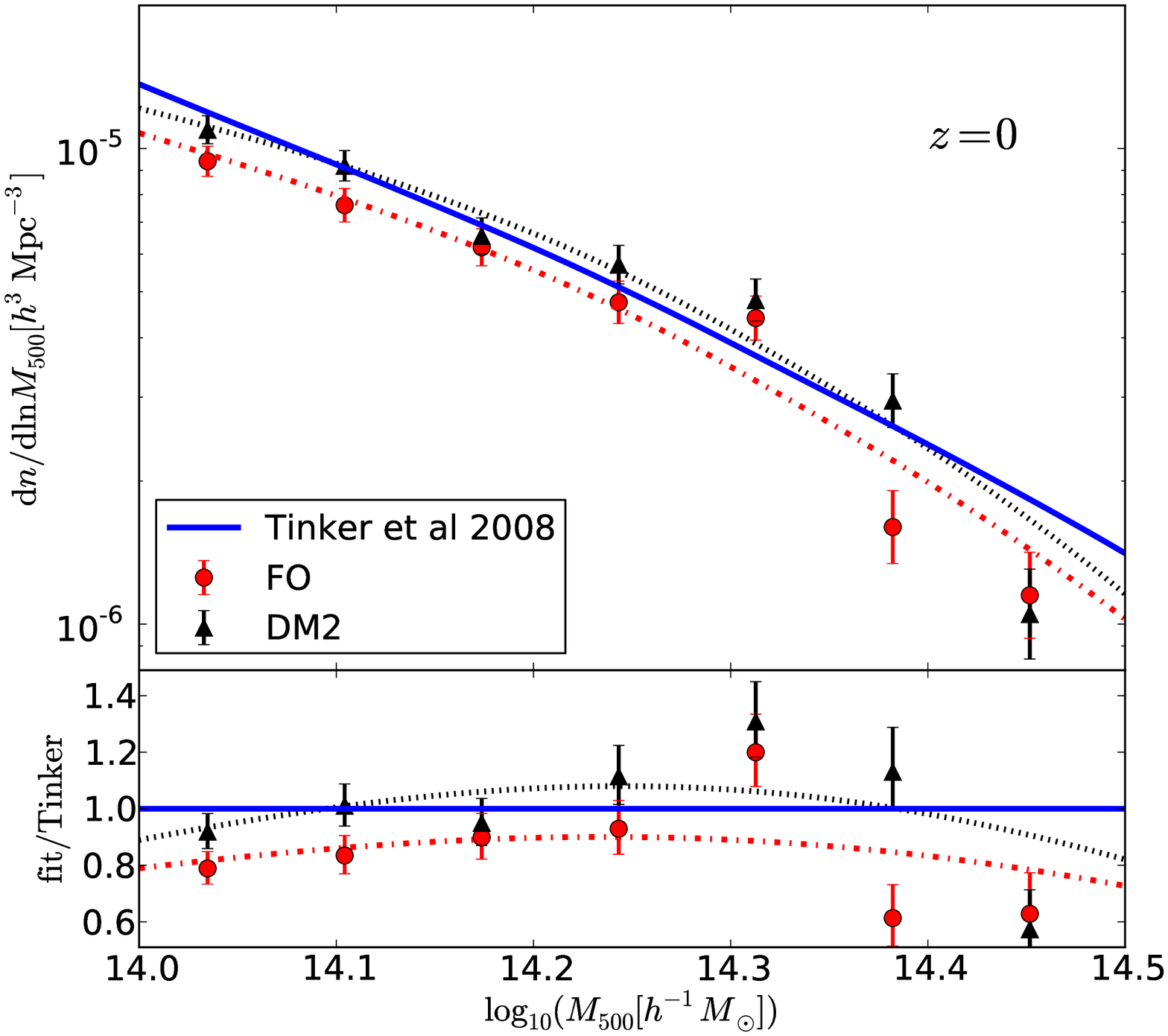}
\caption{Top: Differential mass functions plotted from the two generations of MGS. In the left panel the \GO and \PC results are shown as cyan squares and green diamonds respectively. 
In the right hand panel the \FO and \DMtwo mass functions are plotted as red circles and black triangles respectively. 
Also plotted are fits to each cluster population computed by allowing the parameters in equation \eqref{eq:tinker_function} to vary. 
In each panel the TMF is also shown in solid blue. Lower: Ratio of each mass function best fit with the TMF. 
Over the range plotted here the mean fit/Tinker values are $\simeq 0.82, 0.86$ for the \PC and \FO mass functions respectively.
For a similar comparison of \GO and \PC (using $\Delta=200$) see Fig. 4 of~\citet{Stanek:2008am}.}
\label{fig:MF}
\end{figure*}
Following~\citet{Jenkins2001}, we express the halo mass function as the logarithmic derivative of the number density, $n(M_{500})$, with respect to mass
\beq\label{eq:dndM}
\frac{\mathrm{d}n}{\mathrm{d}\ln M_{500}} = f(\sigma) \bar{\rho}_{\rm m}(z)\frac{\mathrm{d}\ln \sigma^{-1}}{\mathrm{d}M_{500}},
\eeq
where the variance of the density field within spheres of radius $R$ [$h^{-1}$ Mpc], 
\beq
\sigma^2(R,z) = \frac{D^2(z)}{2\pi^2} \int P(k) W^2(kR) k^2 \mathrm{d}k,
\eeq $P(k)$ is the linear matter power spectrum, $D(z)$ is the linear growth factor and $W(x)=3\left(\sin x - x \cos x \right)/x^3$ is the Fourier transform of the real-space top-hat filter.
The function $f(\sigma)$ is independent of cosmological parameters by design.
Recent studies,~\citep{Tinker2008,Watson2012}, have taken the parameterisation
\beq
f(\sigma) = A\left[\left({\beta \over \sigma} \right)^{\alpha} +1\right] \exp \left(-{c \over \sigma^2}\right),
\label{eq:tinker_function}
\eeq
and computing the constants $A,\alpha, \beta, c$ from their respective dark matter-only cosmological simulations for a range of $z$ and $\Delta$.

Fig.~\ref{fig:MF} shows mass functions computed from the cluster distributions in the MGS at $z=0$, where the mass bins were spaced with $\Delta \ln M_{500}=0.16$ and the position of each bin was taken to be the midpoint\footnote{We have confirmed that the conclusions of this paper are not sensitive to either taking the bin midpoint, rather than the mean or median, nor the logarithmic width of the bins.}.
We also plot the appropriate TMF in both panels, where $P(k)$ was calculated using the publicly available code CAMB\footnote{\url{http://camb.info}}~\citep{Lewis2000}.
It is evident from the right hand panel that the TMF is broadly consistent with the dark matter-only simulation for $14 \approxlt \log_{10}\left(M_{500} [\hmsun]\right) \approxlt 14.5$. 
Above this mass, Poisson noise due to rare objects starts to become significant.
The agreement between the \DMone and \DMtwo mass functions and the TMF is within the 5 per cent statistical errors of the~\citeauthor{Tinker2008} mass function fitting at the low mass end of the mass function.
For confirmation of the agreement between \DMone/\DMtwo and the TMF, see Fig.~\ref{fig:contours}.

\subsection{Impact of baryons on mass function}

As discussed in~\citet{Stanek:2008am}, the clusters in the \PC simulation showed a systematic suppression relative to both the \GO clusters and the TMF.
We demonstrate this effect again in the left hand panel of Fig.~\ref{fig:MF} for $\Delta=500$.
One can also see from the right hand panel of Fig.~\ref{fig:MF} that the mass function computed from \FO is also offset from both the data of \DMtwo and the TMF. 

We note that there is agreement between the \GO mass function and the TMF.
In this model the mass independent baryonic depletion, detailed in Section~\ref{sec:bary_dep}, is sufficiently mild that the expected underlying dark matter-only (\DMone) mass function is recovered.
In both the \PC and \FO mass functions, the larger relative offset is a consequence of the lower, mass dependent, cluster baryon fraction resulting from gas ejection processes.

One can parameterise the deviation from the TMF by generalising the mass bias parameter described in Section~\ref{sec:intro} to include the effects effects of baryon depletion.
Since we know the true masses of our simulated clusters, we will ignore the complexities of hydrostatic bias in cluster observations.
Instead we define the baryonic depletion bias $b_{\rm dep}=1-M_{\rm DM}/M_{\rm hyd}$ where $M_{\rm DM}$ and $M_{\rm hyd}$ are the masses of a cluster in dark matter-only and hydrodynamic simulations.
We found that rescaling the mass variable, $M_{500}$, in the \DMtwo mass function at $z=0$ ($z=0.17$) by $1-b_{\rm dep}=0.9$ (0.93) brought the \FO and \DMtwo curves into closer agreement.
We note however, a noticeable difference in shape between the \FO and adjusted \DMtwo mass functions was evident.
While the zeroth order effect of the baryons on the mass function is to shift it relative to the dark matter-only mass function by $10$ per cent ($7$ per cent) we argue that it is insufficient for precision cosmology.
Further, arguments of this type do not account for changes in $R_{500}$ in a consistent manner.

We will return to the problem of baryonic influence on the mass function in Section~\ref{sec:correcting} and discuss the effect of back-reaction of baryons on the dark matter.
It will be shown that the primary reason for the change in mass is ejection of baryons from clusters.

\subsection{Effect on cosmological parameters}
\begin{figure}
\centering
\includegraphics[width=8.5cm]{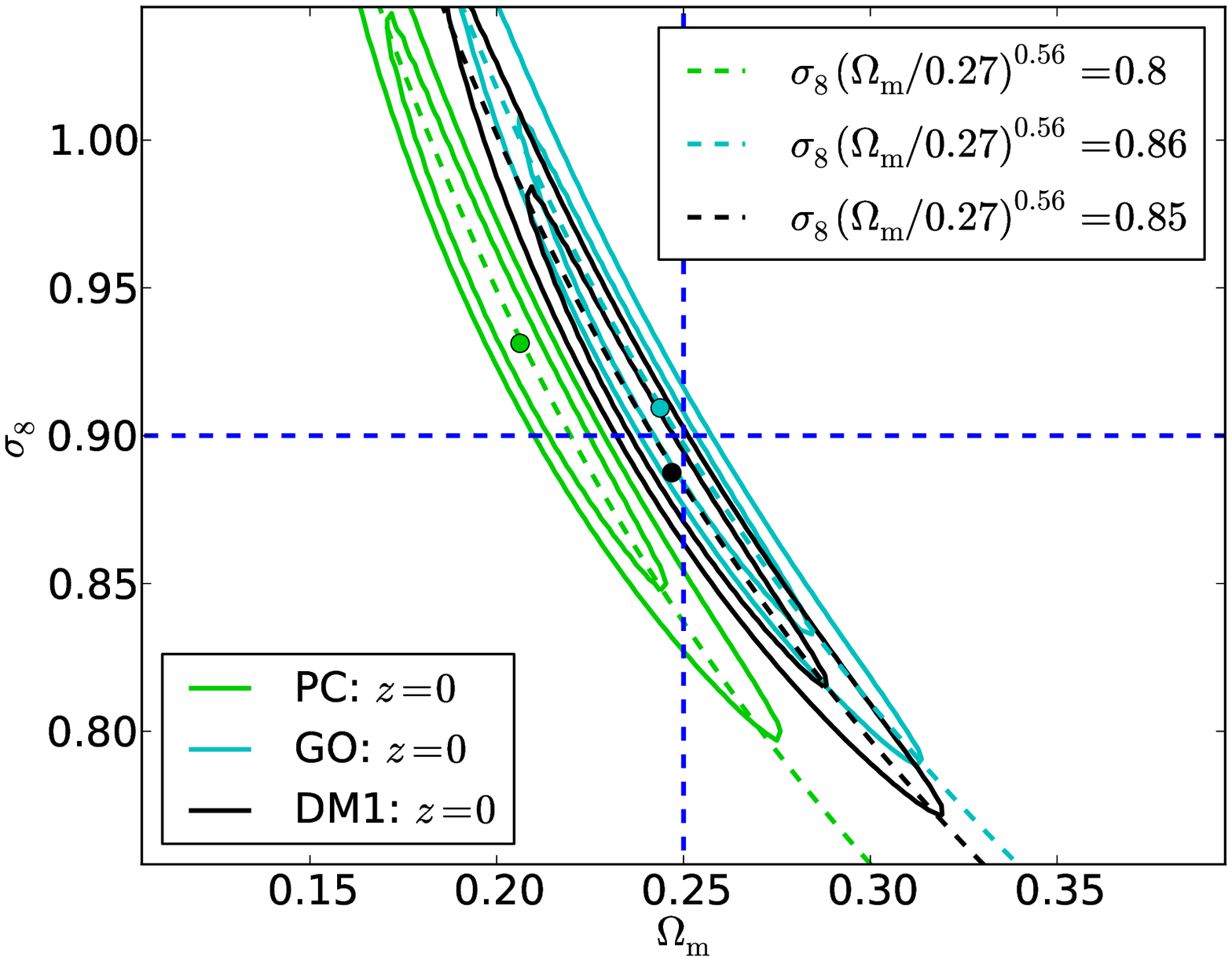}
\includegraphics[width=8.5cm]{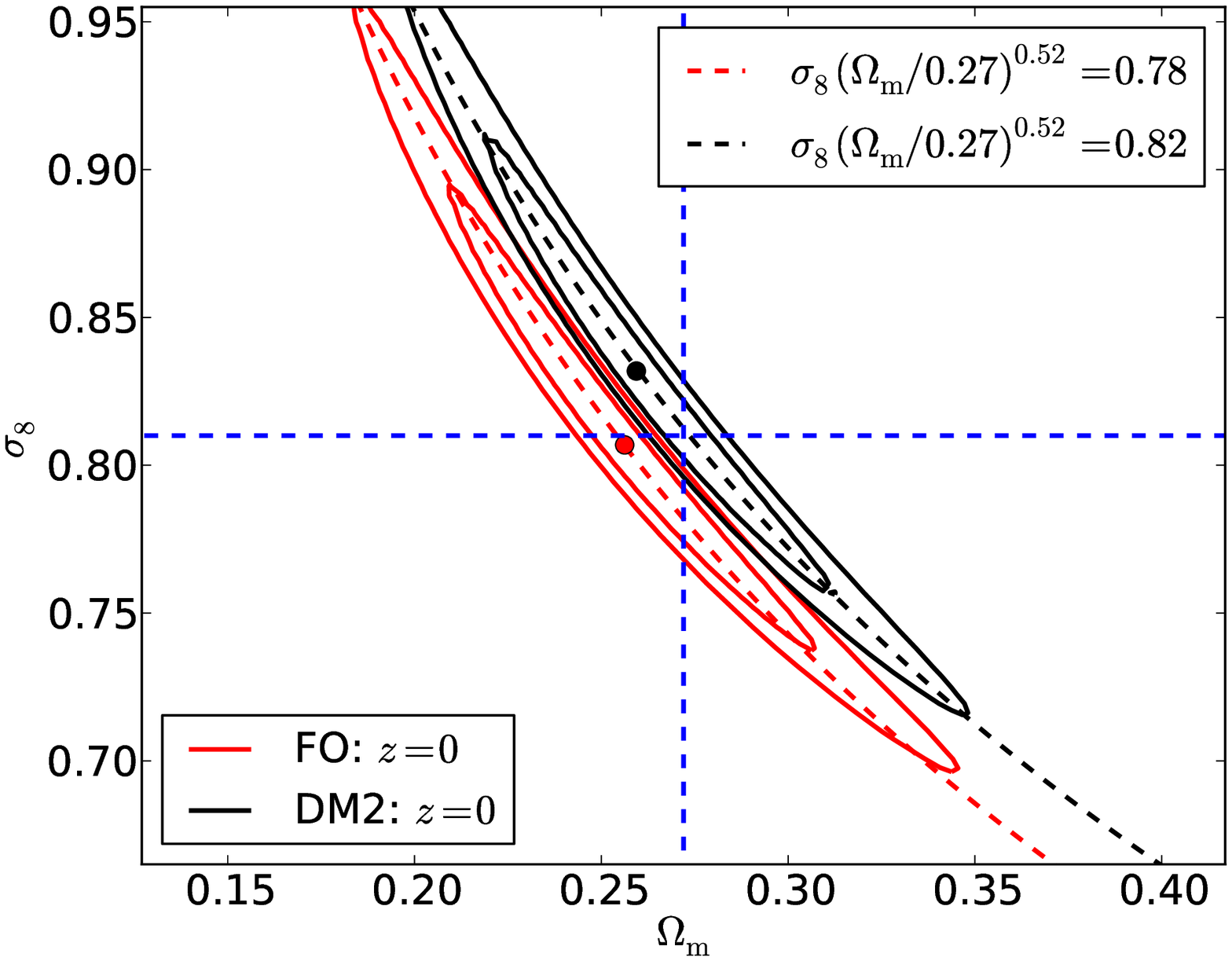}
\includegraphics[width=8.5cm]{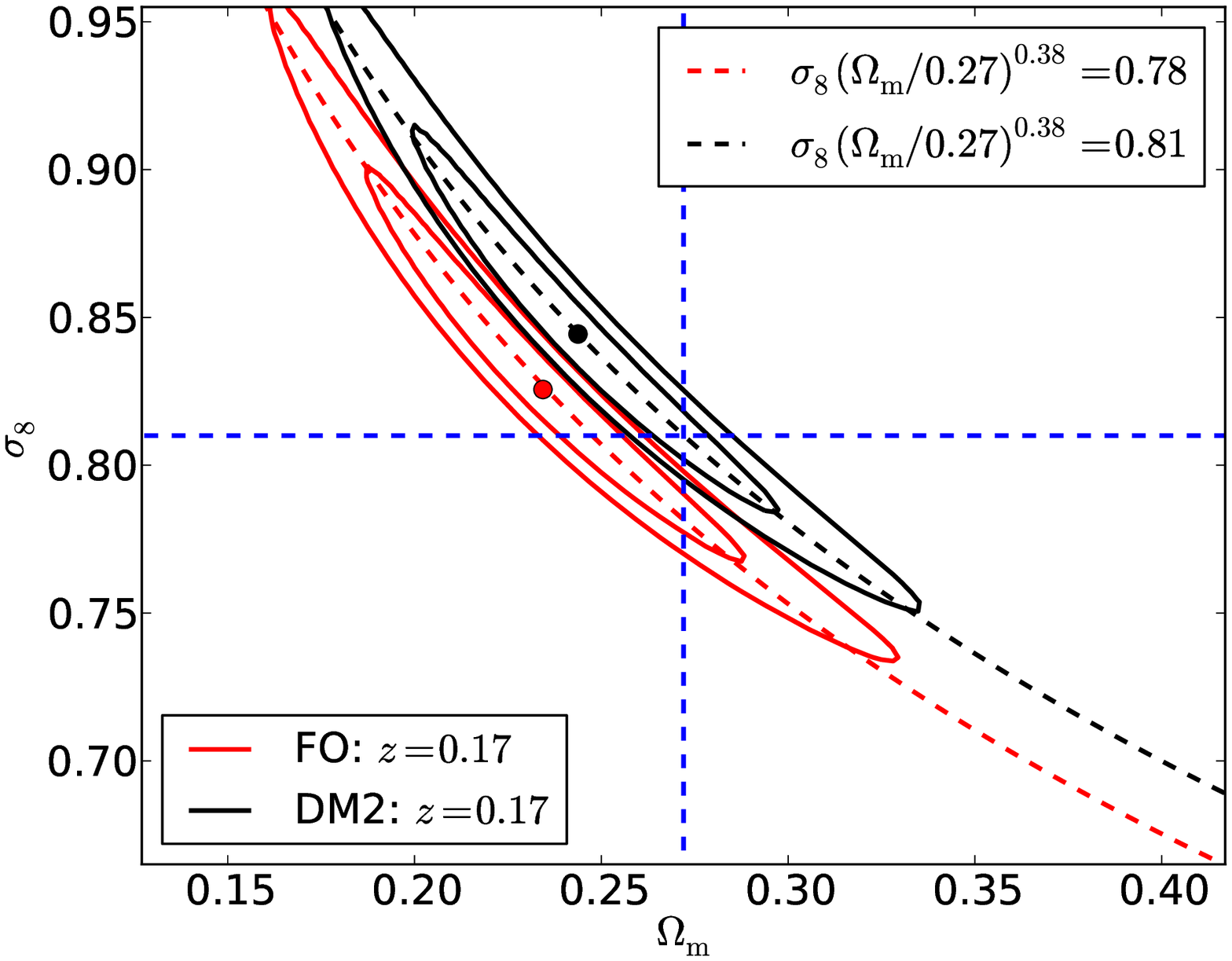}
\caption{Likelihood contours computed from the simulated cluster mass functions at $z=0$ (top and middle) and $z=0.17$ (lower) assuming the TMF (see Fig.~\ref{fig:MF}). In the top panel, cyan contours were computed using the \GO mass function and the green contours were calculated using the results of the \PC simulation. Similarly in the middle and lower panels, the \DMtwo and \FO likelihood contours are shown in black and red respectively. Also shown are the lines of best fit describing the degeneracy between $\omm$ and $\sigma_8$ for each generation. In order to directly evaluate the shift in the degeneracy we enforce the \PC and \GO power-law indices to be that of the \DMone and the \FO power-law index to be that of the \DMtwo. The points of maximum likelihood are shown as coloured dots. The discrepancy between the \PC and \FO distributions and the fiducial values of $\omm$ and $\sigma_8$ (indicated by the blue dashed lines) is the key result of our investigations.}
\label{fig:contours}
\end{figure}

As outlined in the introductory section, the primary function of the TMF is to link measurements of the halo mass function to the $\omm$ and $\sigma_8$ parameters of the underlying cosmology.
By using the TMF to constrain cosmology from galaxy cluster measurements it is assumed that is that the gas content of clusters traces the dark matter component.
Given that we have demonstrated that simulating the baryonic content within clusters suppresses cluster abundance at fixed mass relative to the dark matter-only result (particularly in the \PC and \FO cases), we now investigate the impact of this result on estimations of cosmological parameters.

We use the simulated cluster mass functions described in the previous sections (see Fig~\ref{fig:MF}) as our mock data.
Taking the TMF as our assumed model, we computed likelihood distributions for each of the simulated populations on a regular grid.

At each point in $\omm, \sigma_8$ space the Cash statistic~\citep{Cash1979}
\beq
\begin{split}
C &=-2 \sum_{k=1}^{N_{\rm tot}} \ln \mathcal{P}(N_k|n_k) \\
&= -2 \sum_{k=1}^{N_{\rm tot}} \left(N_k \ln(n_k) - n_k - \ln(N_k!) \right),
\end{split}
\eeq
was calculated, where $\mathcal{P}(N_k|n_k)$ is the probability of finding $N_k$ clusters in a bin given a number $n_k$ predicted by the model.
We then used the fact that $\Delta C$ is distributed as $\Delta\chi^2$ with two degrees of freedom~\citep{Press1992}.

As in Section~\ref{sec:cl_MF}, we used CAMB to calculate $P(k)$ and hence the model $\dndlnM$ through equation \eqref{eq:dndM}.
Throughout we assumed the values of other cosmological parameters were known since they do not contribute significantly to the variance in the mass function measurement \citep{Murray2013}.

The likelihood contours calculated using the MGS clusters are shown in Fig.~\ref{fig:contours}.
As before, in this analysis we conservatively used clusters with $14<\log_{10} \left(M_{500} [\hmsun]\right)<14.5$.
We show the earlier epoch since it is the median $z$ of the 2013 \textit{Planck} SZ high S/N catalogue~\citep{PlanckCollaboration2013e}.
The lines of degeneracy between $\sigma_8$ and $\omm$ are also shown in Fig.~\ref{fig:contours}.

In the upper panel of Fig.~\ref{fig:contours} the offset between the \GO and \PC likelihood contours is clear.
While the peak of the $z=0$ \PC distribution is offset along the degeneracy, the movement of the degeneracy itself is the crucial characteristic.

The likelihood distribution contours resulting from the \FO and \DMtwo mass functions are shown in right hand panel of Fig.~\ref{fig:contours}.
The change in the \FO/\DMtwo power-law indices of the degeneracy between the middle and bottom panels is due to the redshift dependence of the mass function.
As the mass function evolves with redshift, it enables one, in principle, to break the $\omm-\sigma_8$ degeneracy with multi-redshift observations.
The \FO contours are clearly shifted relative to the dark matter-only simulation at both epochs.
We quantify this shift as $\Delta\left[\sigma_8 \left(\omm/0.27\right)^{\gamma_z} \right] = \Delta B_z$, where $\gamma_z$ and $\Delta B_z$ are constants, by fitting a power-law relation to the likelihood contours.
Over the redshifts of interest ($z=0 \to 0.17$) $\gamma_z$ varies from 0.52 to 0.38 whereas $\Delta B_z$ remains $\simeq -0.03$.
The discrepancy between the \textit{Planck} CMB and cluster count measurements can be described as $\Delta\left[\sigma_8 \left(\omm/0.27\right)^{0.3} \right] \simeq -0.08$.
It should be noted that the widths of the error contours (but crucially not the offset $\Delta B_z$) in Fig.~\ref{fig:contours} reflect the size of the simulation volume (and hence the number of clusters) rather than any particular observational survey.
Though we urge caution when comparing simulated snapshots and observations, our calculations demonstrate that the effects of baryonic depletion in clusters is non-negligible in this context. 

\subsection{Correcting for baryonic physics in galaxy clusters}\label{sec:correcting}
\begin{figure}
\centering
\includegraphics[width=8.5cm]{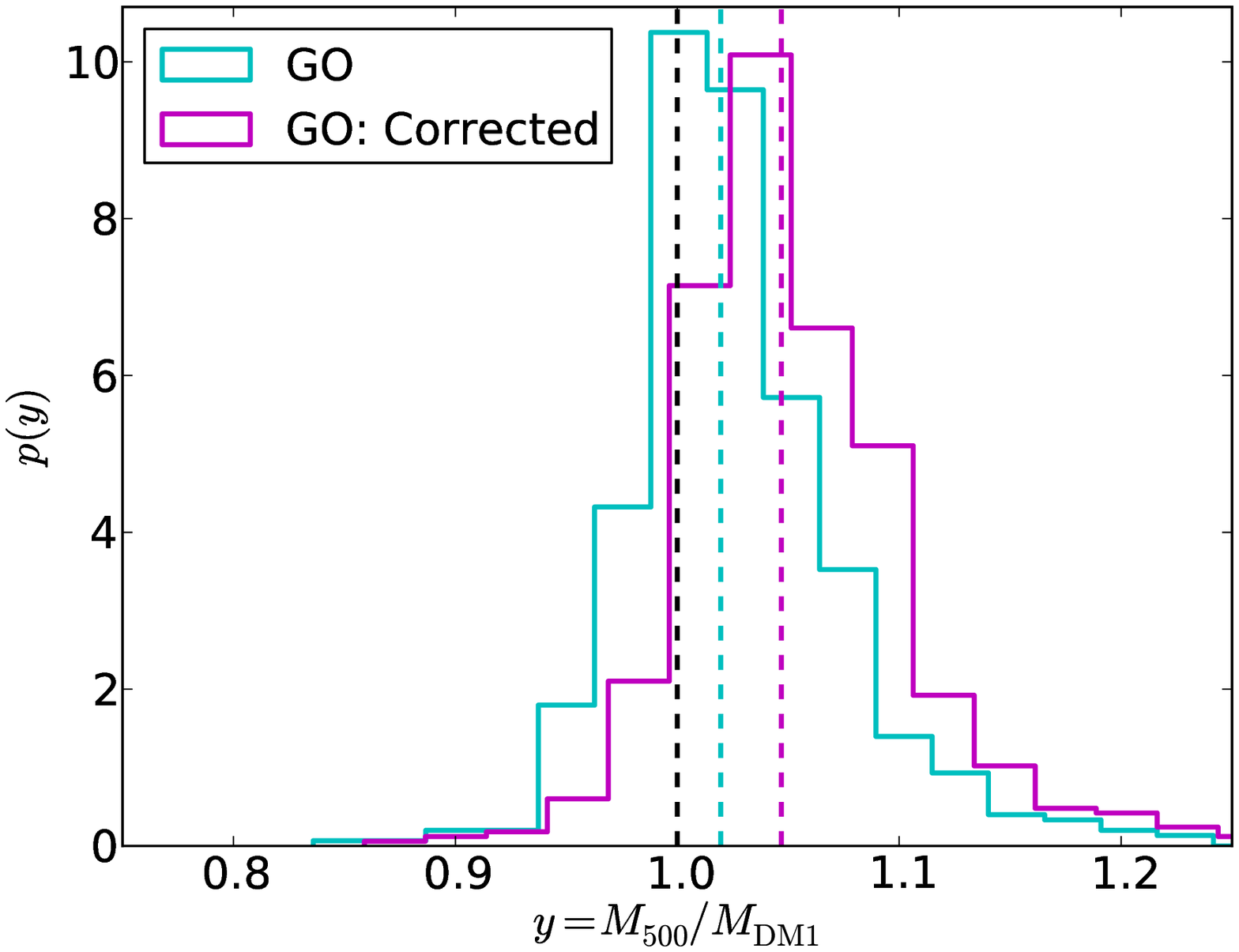}
\includegraphics[width=8.5cm]{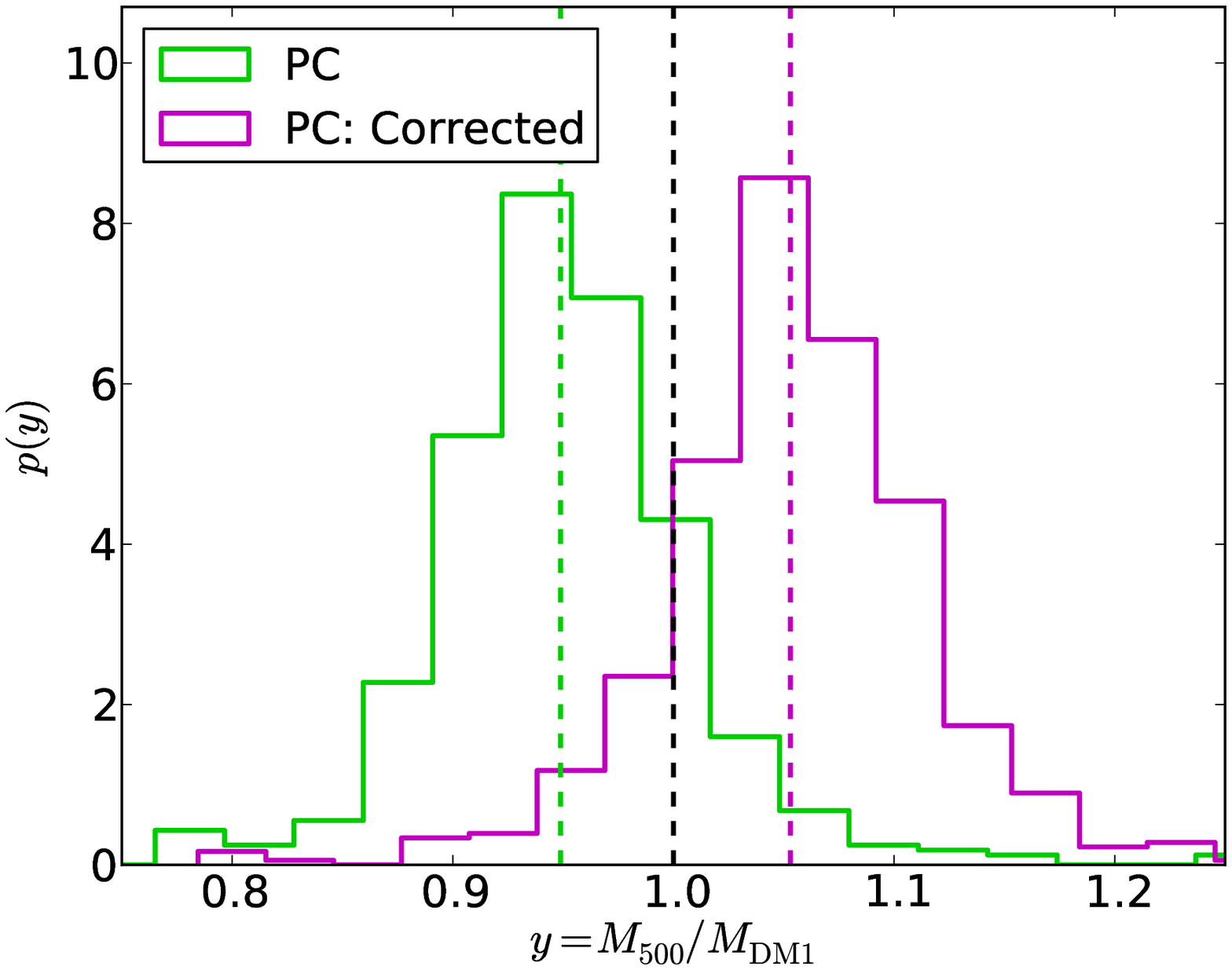}
\includegraphics[width=8.5cm]{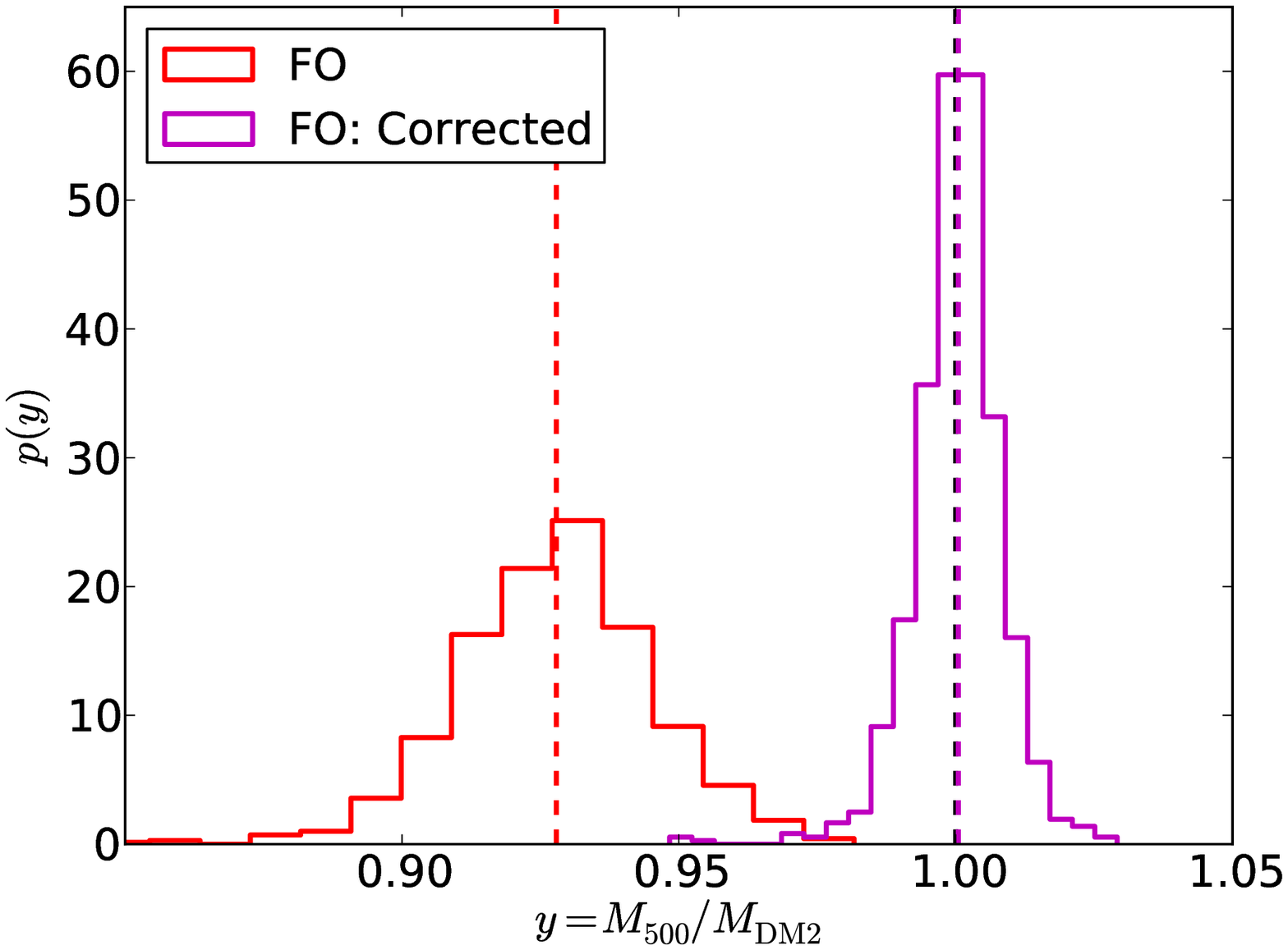}
\caption{Distributions of total mass ratios, $y=M_{500}/M_{\rm DM1/2}$, where $M_{500}$ is the total mass of a cluster in the hydrodynamical simulation (\GO, \PC and \FO are shown in cyan, green and red respectively) and $M_{\rm DM1/2}$ is the mass of the same cluster in the appropriate dark matter-only simulation. 
We also plot the ratio of the cluster masses after correcting the hydrodynamical mass as outlined in Section \ref{sec:correcting}. 
In the \FO model, the correction is near exact by construction; the $\pm0.006$ scatter about $y=1.001$ is due to numerical error in recomputing $R_{500}$ from the $M_{\rm est}(<r)$ profile rather than the particle distribution. 
In the GO and \PC models, where baryons can influence the dark matter density profile, the corrected halo masses do not exactly match the cluster masses in \DMone.}
\label{fig:hist_corr}
\end{figure}

As we have demonstrated in the previous sections, the mass of a given cluster in a hydrodynamic simulation is not equal to the mass of the same object in a dark matter-only simulation.
We now outline and test a method for ``correcting'' a baryon influenced cluster mass function in order to enable one to use the TMF (or similar) for cosmological parameter determinations.

Our proposed three-step methodology is as follows:
\begin{itemize}
\item Calculate the \textit{dark matter} mass profile of each cluster, $M(<r)$, using knowledge of the total density profile and removing the stellar and gas components;
\item Supplement the dark matter mass with baryons such that the baryon fraction is equal to the cosmic value everywhere in the cluster, i.e.\ estimate the mass profile $M_{\rm est}(<r)=M_{\rm DM}(<r)/(1-\omb/\omm)$;
\item Recalculate $R_{500}$ and $M_{{\rm est},500}$ using the new mass profile.
\end{itemize}

Observationally, it is too expensive to calculate the dark matter mass profile of every cluster in a survey. This would require, for example,
high quality X-ray data allowing the estimation of total density and temperature profiles, or weak lensing data with sufficiently 
high density of background sources. Additionally, the mass distribution of gas (using X-ray data) and stars (including any additional
diffuse component) would also be required. 
In practice, a mass-observable relation (ideally with minimal scatter) is calibrated for a smaller number of clusters and the observable used as a mass {\it proxy} for the full sample (e.g.~\citealt{Arnaud2007}). 
The practice of mass proxy calibration is common to all cluster surveys including the \textit{Planck} analysis~\citep{PlanckCollaboration2013}.
In this case, the simple procedure outlined above could be applied to re-calibrate the mass-observable relation for cosmological purposes (we leave feasibility of such a procedure to future study).

The advantage of this procedure (rather than attempting to correct the theoretical mass function) is that it is empirical, relying on only the observational data and not assuming any theoretical model for how the baryons affect the total cluster mass. 
However, it does assume that the baryonic processes do not significantly influence the underlying dark matter mass profile of clusters.

We test our methodology using the clusters in the \GO, \PC and \FO simulations.
In Fig.~\ref{fig:hist_corr} we plot the ratio of the mass of individual clusters in the hydrodynamical simulations and the dark matter-only simulations.
We also plot the same mass ratio after the correction procedure was applied to the hydrodynamic cluster masses.
The halo structures in \PC were matched with their \DMone counterparts by considering the dark matter-only halos within $0.5 R_{500}$ of the cluster centre.
A match was found for around 98 per cent of the \PC clusters.
In the \FO case, each cluster was mapped directly to the equivalent \DMtwo halo using the data from the \texttt{SUBFIND} analysis of the dark matter distribution. 
In the absence of baryon-dark matter back-reaction we expect the corrected distributions to be centred at unity.

\begin{figure}
\centering
\includegraphics[width=8.5cm]{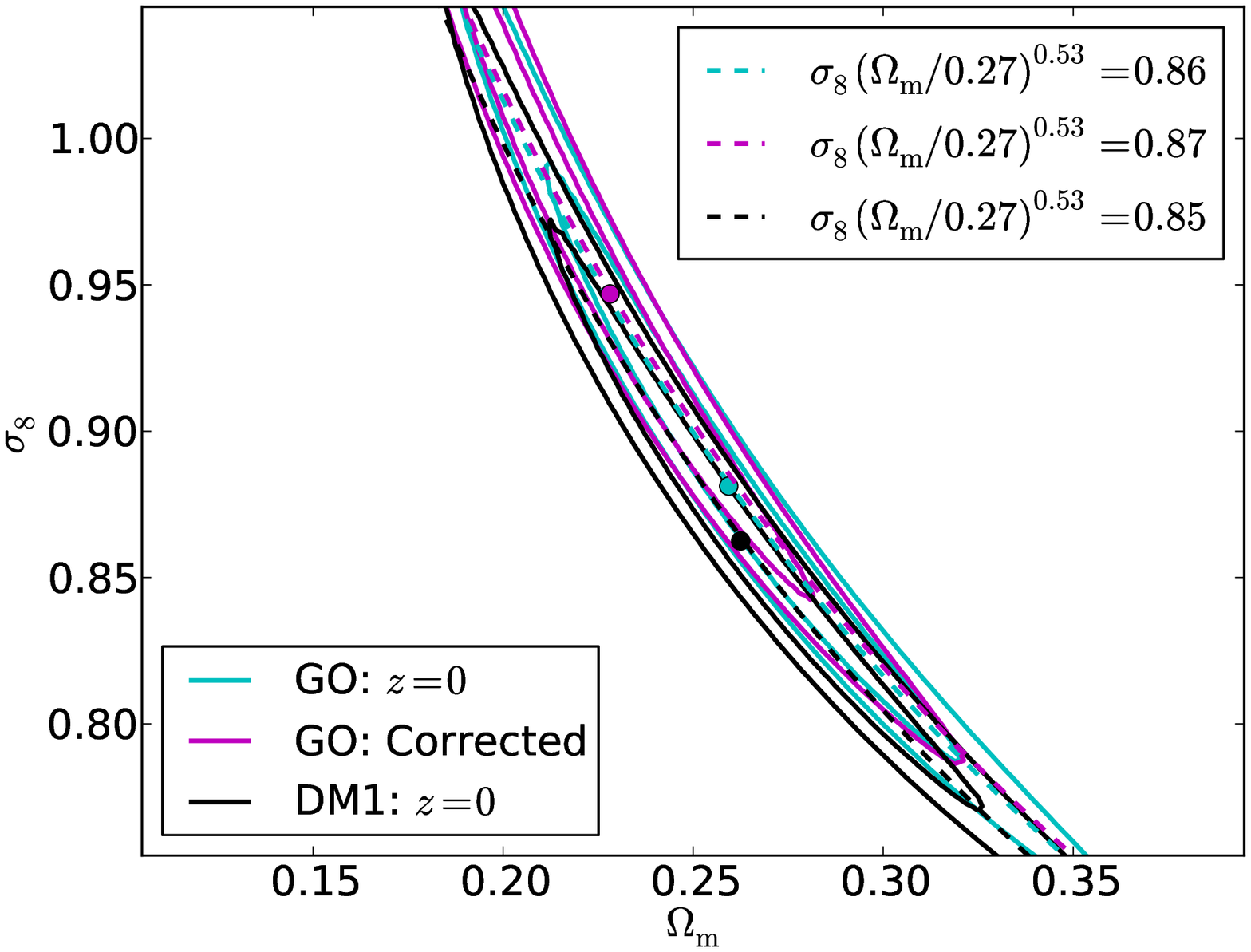}
\includegraphics[width=8.5cm]{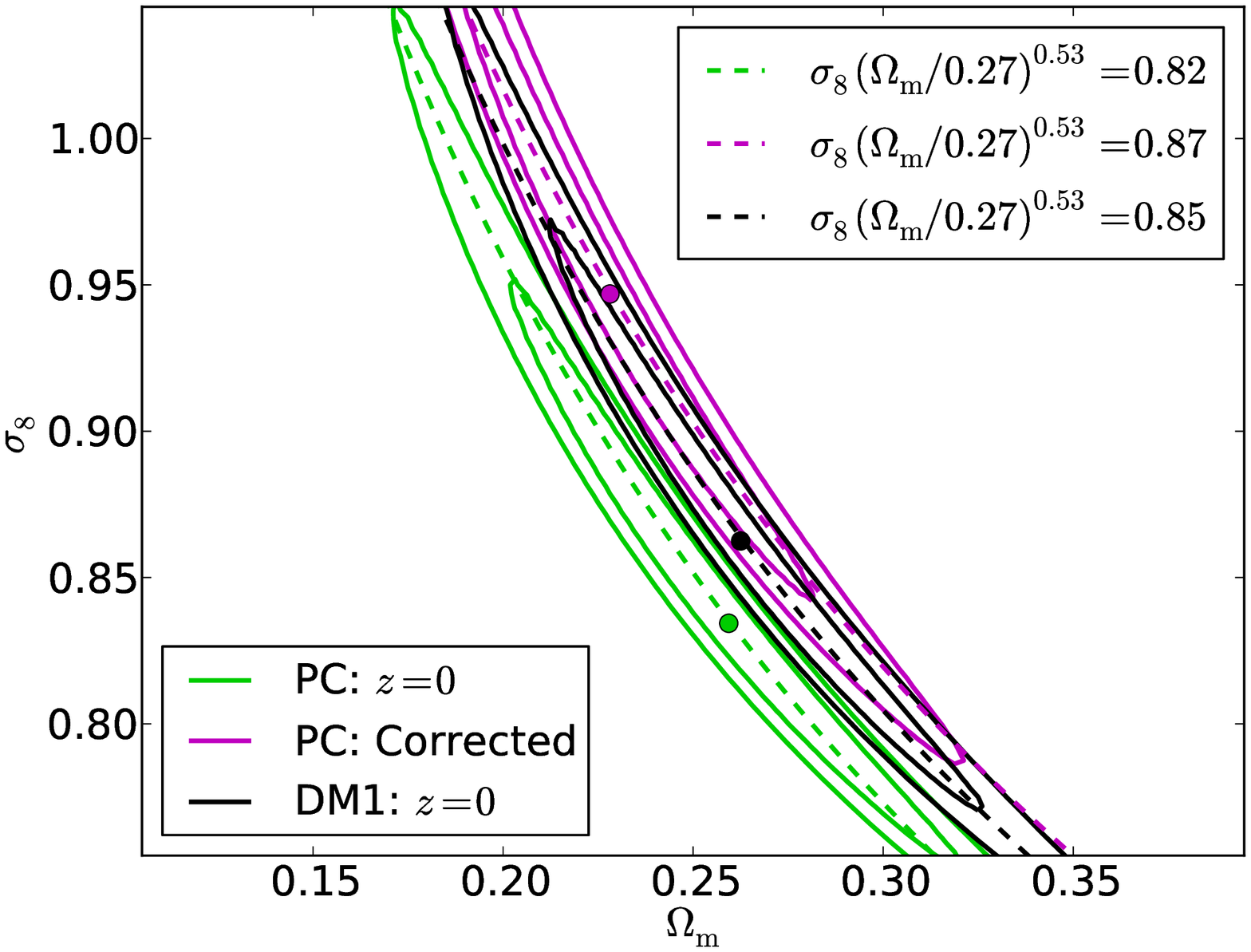}
\includegraphics[width=8.5cm]{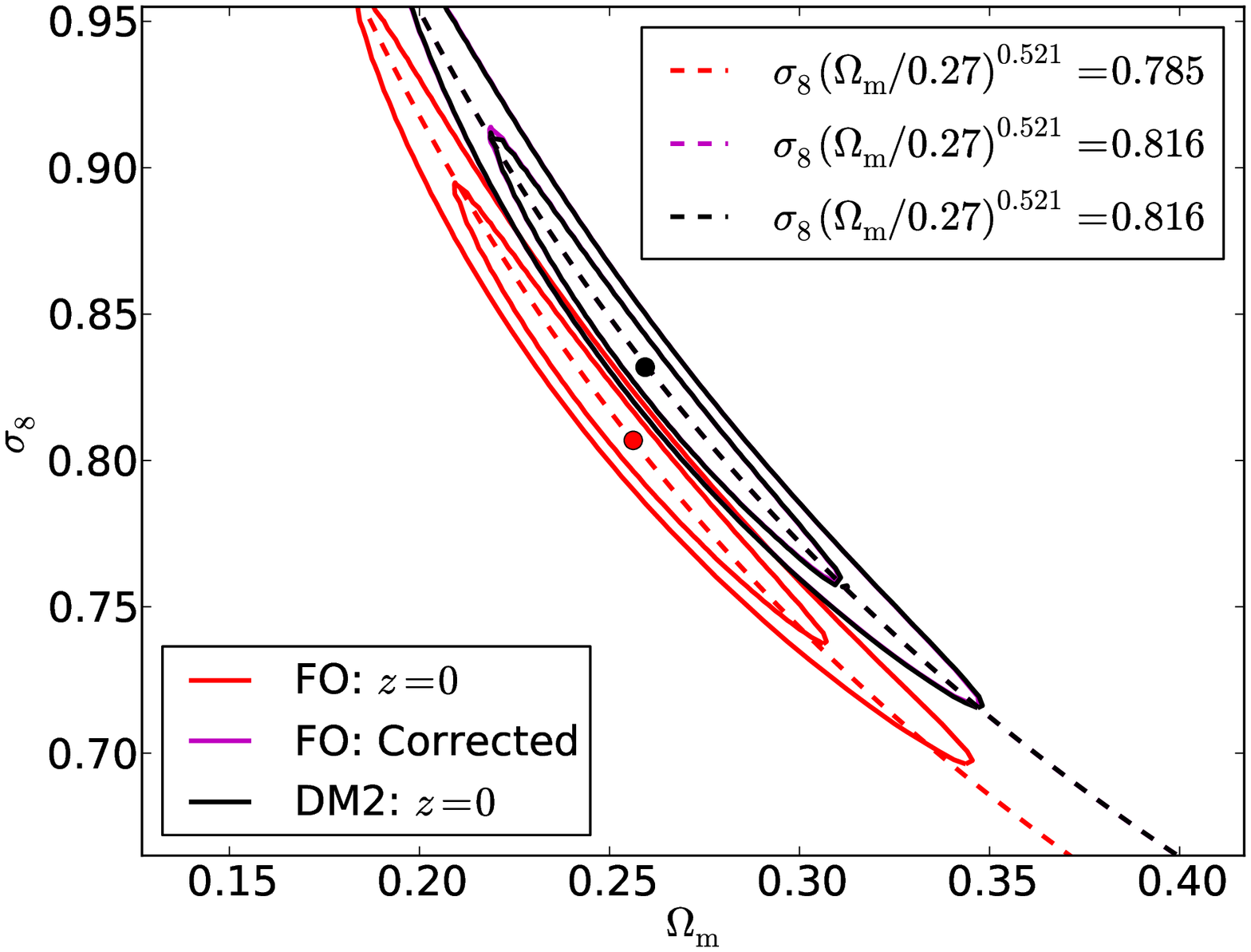}
\caption{Likelihood contours calculated from $z=0$ mass functions in the same manner as those in Fig.~\ref{fig:contours}. ``Corrected'' distributions were derived from the cluster mass function after applying the procedure to each cluster. As before, the results from the \GO, \PC and \FO simulations are shown in cyan, green and red respectively. The likelihood distributions calculated using the corrected mass functions are shown in purple. Additionally, the distributions obtained using the \DMone/ \DMtwo simulations are shown in black. The power-law index of the fitting is fixed to the value of the dark matter-only degeneracy}
\label{fig:contours_corr}
\end{figure}

In the \FO simulation, the dark matter particles are not gravitationally influenced by the baryons and so no back-reaction is possible \citep{Short:2012ck}.
This fact is demonstrated by the purple distribution in the right hand panel of Fig.~\ref{fig:hist_corr} being centred around unity.
The small amount of scatter is due to the error in calculating $M_{500}$ from cluster profiles rather than the actual particles themselves.
As a further test of our method, we confirm in the right hand panel of Fig.~\ref{fig:contours_corr} that the corrected \FO mass function reproduces the \DMtwo mass function.

The \GO and \PC implementations, on the other hand, do evolve the dark matter and gas consistently.
As shown in the upper and middle panels of Fig.~\ref{fig:hist_corr}, the median mass ratios for the \GO and \PC clusters are 1.046 ($\protect\sigma_{\rm \GO}=\begin{smallmatrix}+0.039 \\ -0.052\end{smallmatrix}$) and 0.949 ($\protect\sigma_{\rm \PC}=\begin{smallmatrix}+0.048 \\ -0.045\end{smallmatrix}$) respectively; whereas after correction procedure the median values are 1.020 and 1.053 with scatter $\protect\sigma_{\rm \GO}=\begin{smallmatrix}+0.033 \\ -0.048\end{smallmatrix}$ and $\protect\sigma_{\rm \PC}=\begin{smallmatrix}+0.047 \\ -0.054\end{smallmatrix}$.
The quoted values of scatter were calculated from the 16th and 84th percentiles of the distributions.
The fact that the corrected distributions are not centred on unity with minimal scatter is a reflection of the baryon-dark matter back-reaction\footnote{We also checked that at lower overdensity ($\Delta=200$), the difference between \DMone and \GO masses was smaller than at $\Delta=500$.}. This irreducible effect is the central limitation of our method. Without a greater understanding of the effects of baryons on the gravitational potential, the procedure may be unable to recover the abundance of clusters to within around $5$ per cent.

In Fig.~\ref{fig:contours_corr} we compute likelihood contours from the \GO, \PC and \FO mass functions before and after applying the above correction procedure. 
Note that because of the mass limit made on the \PC cluster catalogue and the fact that the correction invariably increased the mass of clusters, we only consider clusters with $\log_{10}\left(M_{500} / \hmsun\right)>14.1$.
This additional condition decreases the number of clusters in the mass function and therefore increases the width of the contours shown in the left hand panel of Fig.~\ref{fig:contours_corr} as well as moving the contours along the $\sigma_8-\omm$ degeneracy.
In the lower panel of Fig.~\ref{fig:contours_corr}, the corrected \FO contours map almost directly onto the \DMtwo contours.
This excellent agreement is due to the fact that, in the \FO model, baryons do not influence the dark matter mass profile: scaling the dark matter mass profile by $\left(1-\omb/\omm \right)^{-1}$ recovers the \DMtwo mass distribution by design.

As expected from the offset in the \GO/\DMone and \PC/\DMone cluster mass ratio distributions shown in Fig.~\ref{fig:hist_corr}, the contours from the corrected mass functions do not match the \DMone contours. 
In fact, the correction procedure does not appear to improve agreement between the \GO/\PC contours and the \DMone. 
The fact that the input cosmology is not recovered by this correction procedure is a demonstration that the baryons do have a significant influence on the shape of the dark matter mass profile in these models.

\section{Discussion}\label{sec:discussion}

In this paper we have discussed the impact that baryonic physics can have on the observed cluster mass function.
Although further study is required in order to fully model the gas physics in clusters, we have shown that the baryon fraction measured in MGS is broadly consistent with observations. 
We argue that since the baryon fraction is similar to that observed in other simulated clusters~\citep{Sembolini2013,Planelles2013}, the suppression in the mass function shown in Fig.~\ref{fig:MF} is generic to realistic baryonic treatments.

In contrast to our findings, a $\sim 7$ per cent overabundance of clusters relative to dark matter only haloes was reported by~\citet{Cui2012a}.
In their simulations efficient radiative cooling of gas ensures the hydrodynamically simulated clusters are more concentrated than their dark matter-only counterparts and therefore have larger values of $M_{500}$.
We reason that the lack of AGN feedback or early heating in their simulations allowed clusters to retain their baryon content and thereby allowed the mass of a given cluster to increase relative to its dark matter-only counterpart.
Due to the functional form of the halo mass function, a shift in mass of this nature would result in an increase in the number of clusters of a fixed mass.
As argued in~\citet{VanDaalen2011}, a realistic treatment of gas dynamics results in a decrease in the matter power spectrum relative to dark matter only simulations on scales $1\approxlt k \approxlt 10~h \mathrm{Mpc}^{-1}$.
It therefore follows that a relative underabundance of clusters would be expected.

We have shown that assuming the TMF leads to an incorrect measurement of the $\sigma_8-\omm$ degeneracy by $\Delta\left[\sigma_8\left(\omm/0.27\right)^{0.38}\right]\simeq -0.03$ at $z=0.17$ when considering realistic clusters with $14<\log_{10}\left(M_{500} [\hmsun]\right)<14.5$.
The discrepancy we describe here is not specific to the TMF.
We confirmed that using the~\citet{Watson2012} fit as the assumed model, instead of the TMF, produced a similar offset between the derived dark matter-only and hydrodynamic likelihood distributions.

The analysis of~\citet{Balaguera-Antolinez2013} came to similar conclusions as presented in this work, though through different means.
In that study, predictions regarding the observed mass function were made using the TMF and assuming the form of $\fb(M_{500})$ from~\citet{Lagana2011}.
By creating mock catalogues, the above authors showed a systematic shift of the similar order and sense as that shown in Fig~\ref{fig:contours}.
Recently however, \citet{Martizzi2013} extended the~\citeauthor{Balaguera-Antolinez2013} methodology using the $f_{\mathrm{b}}-M_{500}$ relation derived from their set of high resolution cluster resimulations.
They concluded that the mass function should be boosted by the effects of baryonic physics because, in contrast with the observational data shown in Fig.~\ref{fig:baryon_fraction}, the baryon fraction they use is higher than the cosmic mean over the mass range.
Our investigations differ from the above methodology in that we make no assumption regarding the functional form of the cosmic mass function or the $\fb-M_{500}$ relation at the run-time of our simulations.
Further, we have shown the $\omm-\sigma_8$ degeneracy offset is present in two very physically distinct scenarios (\PC and \FO).

In future \textit{Planck} analyses, where lower mass clusters are studied, the influence of baryonic physics on cluster masses will have to be considered and accounted for.
Further, our calculations are applicable to any cosmological survey with clusters of mass $\approxlt 10^{14.5} \hmsun$ (particularly XCS) and provide qualitative information on what might happen at larger masses, although we leave this for future investigations. 

In our final section we proposed and tested a model-independent procedure designed to recover the results of dark matter-only simulations from measurements of clusters with baryonic effects.
The correction procedure was demonstrated to work well in the case where baryon-dark matter gravitational interaction was neglected (\FO model).
However in simulations in which baryons significantly contribute to the gravitational potential (\GO/\PC), the procedure was deemed to be insufficient.
We conclude that further modelling of baryonic physics in clusters is required in order to ensure that future cluster surveys are able to make unbiased constraints on cosmological parameters.

\section*{Acknowledgments}
We thank the referee, Joop Schaye, for their very helpful comments which have improved the quality of the manuscript.
We also thank Frazer Pearce for providing the data from the \DMone simulation.
This work used the DiRAC Data Centric system at Durham University,
operated by the Institute for Computational Cosmology on behalf of the STFC DiRAC HPC Facility (\url{www.dirac.ac.uk}). This equipment was funded by BIS National E-infrastructure capital grant ST/K00042X/1, STFC capital grant
ST/H008519/1, and STFC DiRAC Operations grant ST/K003267/1 and Durham
University. DiRAC is part of the National E-Infrastructure.
In addition, some of the simulations used in this paper were performed at the University of Nottingham HPC Facility.
SJC is supported by an STFC quota studentship and the University of Manchester President's Doctorial Scholar Award.
PAT acknowledges support from the Science and Technology Facilities Council (grant number ST/I000976/1).

\bibliographystyle{mn2e}


\makeatletter
\def\thebiblio#1{%
 \list{}{\usecounter{dummy}%
         \labelwidth\z@
         \leftmargin 1.5em
         \itemsep \z@
         \itemindent-\leftmargin}
 \reset@font\small
 \parindent\z@
 \parskip\z@ plus .1pt\relax
 \def\newblock{\hskip .11em plus .33em minus .07em}
 \sloppy\clubpenalty4000\widowpenalty4000
 \sfcode`\.=1000\relax
}
\let\endthebiblio=\endlist
\makeatother

\label{lastpage}

\end{document}